# Spin accumulation at nonmagnetic interface induced by direct Rashba - Edelstein effect

Florent Auvray*, Jorge Puebla[#], Mingran Xu*, Bivas Rana[#], Daisuke Hashizume[#] and Yoshichika Otani*, [#]

*Institute for Solid State Physics, University of Tokyo, Kashiwa, Chiba, 277-8581, Japan

[#]RIKEN Center for Emergent Matter Science, Wako, Saitama, 351-0198, Japan

Corresponding Author:
Jorge Puebla,
e-mail: jorgeluis.pueblanunez@riken.jp

**Rashba effect describes how electrons moving in an electric field experience a momentum dependent magnetic field that couples to the electron angular momentum (spin). This physical phenomenon permits the generation of spin polarization from charge current (Edelstein effect), which leads to the buildup of spin accumulation. Spin accumulation due to Rashba-Edelstein effect has been recently reported to be uniform and oriented in plane, which has been suggested for applications as spin filter device and efficient driving force for magnetization switching. Here, we report the X-ray spectroscopy characterization Rashba interface formed between nonmagnetic metal (Cu, Ag) and oxide ($Bi_2O_3$) at grazing incidence angles. We further discuss the generation of spin accumulation by injection of electrical current at these Rashba interfaces, and its optical detection by time resolved magneto optical Kerr effect. We provide details of our characterization which can be extended to other Rashba type systems beyond those reported here.**

*Spin accumulation, Rashba – Edelstein effect, nonmagnetic interface, Kerr effect*

# Introduction

On demand generation and efficient detection of spin accumulation is a highly ranked topic in Spintronics. More than 10 years ago, the first experimental evidence of the spin Hall effect was achieved by the optical observation of spin accumulation, induced by passing unpolarized current in bulk GaAs and strained InGaAs without external magnetic field [1]. Since then, several reports have confirmed the existence of spin Hall effect in diverse material systems, and more recently, reports have showed the application of spin accumulation as means for magnetization switching [2,3]. Although, the optical observation of spin accumulation via the magneto optical Kerr effect (MOKE) is routinely applied in semiconductor structures, the feasibility of its application for observing spin accumulation in metallic systems have created controversy, and it has been recently under discussion [4-9]. Intrinsically, metallic elements have relatively low optical activity when compared with semiconductors with well-defined optical transitions. This characteristic harms the signal to noise ratio of MOKE measurements in metallic elements. Particularly challenging is the optical detection of spin accumulation with in-plane orientation induced by spin Hall effect. For spin Hall materials the orientation of spin accumulation is perpendicular to the electron flow and opposite at opposite sample planes; while at both lateral ends of the sample there is uniform out of plane spin accumulation across the sample thickness; closer to the center, the orientation of the spin accumulation is in - plane and gradually changes to the opposite in - plane orientation across the sample thickness. Nevertheless, recent reports showed evidence of optical detection of in-plane spin accumulation in metallic spin Hall effect materials. O. M. J. vant Ervea et. al. [4], defocus the laser beam in his longitudinal MOKE measurements, increasing the spot size to collect larger number of spins from the top sample surface, while at the same time minimizes the laser penetration depth and detection of opposite oriented spins from the bottom sample surface. In an independent report, C. Stamm et. al. [8], relies in the ultra-high sensitivity of his longitudinal MOKE system for the optical detection of in-plane spin accumulation, detecting Kerr rotation angles as low as $\mathbf{10^{-9}\ rad}$. C. Stamm and colleagues go further, and provide a DFT (density functional theory) analysis that allows them to estimate the spin hall angles for Pt and W. However, the MOKE detection of spin accumulation in metallic systems have created skepticism, with particular concern in the influence of heating generated by the injection of relatively large electrical current densities [6,7]. Different from spin Hall effect, where spin orientation is opposite at opposite sample planes, orientation due to direct Rashba - Edelstein effect (DREE) is predicted to be uniform, oriented in-plane and perpendicular to the current direction [10, 11]. Here, we report spin accumulation induced by DREE and extend our discussion in our sample characteristics, X-ray spectroscopy, time resolved transverse magneto optical Kerr effect setup and the analysis of our estimated signals when compared with previous reports in spin Hall materials.

## Experimental details

### Fabrication of nonmagnetic interfaces

The samples were fabricated by two step fabrication method. At first, rectangular structures with dimension of $200 \mu m \times 600 \mu m$ were made on a thermally oxidized Si (001) substrate by maskless UV photolithography, followed by deposition of film-stacking structures of Cu 20 nm and $Bi_2O_3$ 10 nm, and Ag 20 nm and $Bi_2O_3$ 10 nm by electron beam evaporation at room temperature at a base pressure of $10^{-7}$ Torr. In the second step, the contacts and bond pads were designed by photolithography followed by deposition of Ti(5 nm)/Au(150 nm) by electron beam evaporation. The thickness of our samples is optimized to minimize the interaction of our testing laser with the thermally oxidized Si (001) substrate. The absence of spatial inversion symmetry at the Cu/$Bi_2O_3$ and Ag/$Bi_2O_3$ interfaces induces spin orbit interaction of the Rashba type, as we will comment further in the next sections.

### Grazing angle and energy-dispersive X-ray spectroscopy

We characterize the crystal structure at our interfaces by a grazing-incidence X-ray diffraction (GI-XRD) technique. GI-XRD limits the penetration depth of X-ray beam. Below a critical incidence angle, an evanescent wave is created for a short distance and is damped exponentially, allowing Bragg reflections to limited thickness. We use a parallel beam configuration with wavelength, $\lambda = 0.154184$ nm (CuK$\alpha$), and divergence, scattering and receiving slits were set to 1 mm each. XRD source voltage and current are 40 kV and 30 mA, respectively. Further confirmation of the chemical elements contained in our samples was carried out by energy dispersive X-ray (EDX) spectroscopy measurements, with Cartesian geometry optical kernel and monochromatic secondary target excitation to enhanced signal to noise ratio.

**Magneto optical Kerr effect**

For the measurement of spin accumulation at Rashba-like interface in our device we use a custom-made benchtop time-resolved transverse magneto-optical Kerr effect (TR-TMOKE) microscope [12, 13], see figure 1.

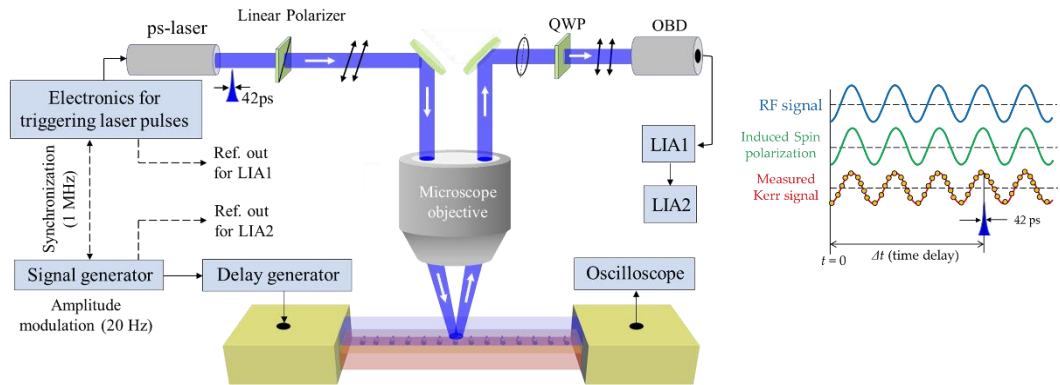

**Figure 1.** Experimental setup for time resolved transverse magneto – optical Kerr effect measurements (TR-TMOKE). The incidence light is polarized at $\eta = 45°$, mixing s and p polarizations. Synchronization of two lock-ins allows to measure the time varying in-plane component of spin polarization.

Two wide bond pads with millimeter dimensions are designed at both ends of the rectangular shaped device for ease of electrical connections. The device is first mounted on a specially designed stage where two semi-rigid coaxial (SMA) cables are connected to the large bond pads with the help of screws. The sample stage is then placed on a x-y-z piezostage. A radio-frequency (RF) signal generator (SG) is connected to one of the SMA cables for applying RF voltage ($V_{RF}$) across the device. Another SMA cable is connected to an oscilloscope to observe the transmitted signal through the device. When $V_{RF}$ is applied across the device, the charge current, passing through the non-magnetic/oxide (Rashba-like) interface, produces nonequilibrium spin accumulation at the interface with in-plane polarization perpendicular to direction of charge current. A pulsed laser beam ($\lambda = 408 nm$) with 42 ps pulse duration and repetition rate of 1 MHz is used for the detection of time varying spin accumulation. In our TR-TMOKE setup, the incidence light is polarized at $\eta = 45°$, mixing s and p polarizations. A systematic study showed an increase of more than two orders of magnitude for detection of MOKE setups by mixing s and p polarization than when working with a well-defined s or p polarization [14]. The laser beam is focused onto the device surface by a long working distance microscope objective (numerical aperture = 0.14, magnification = 5x) with a spot size of about 3.55 $\mu m$ determined by diffraction. In this case, the incident laser beam is slightly shifted from the axis of microscope objective, so that the reflected beam from sample surface can be collected by same microscope objective through diametrically opposite side. This enables us to measure the time varying in-plane component of spin polarization very efficiently. As the Kerr ellipticity of the Rashba interface is stronger than Kerr rotation due to strong dichroism effect [15], we use a quarter wave plate (QWP) to convert Kerr ellipticity into Kerr rotation. The converted Kerr rotation (we call it Kerr signal) of reflected beam which is proportional to the density of spin accumulation is then measured by an optical bridge detector (OBD) in terms of voltage. For the measurement of time varying Kerr signal, the RF current from SG is sent to the devices via an electronic delay generator which helps to introduce variable delay between RF current and laser pulse up to 40 ns with a time resolution of 1 ps. Spin accumulation signals are stroboscopically measured by locally probing sample surface with laser pulses at various time delays with respect to a reference time delay (we call zero-time delay). The reference time delay is decided by the internal delay of electronic devices and the length of SMA cable. The amplitude of the rf signal is modulated by 100% at a very low frequency of 20 Hz. We use two lock-in-amplifiers (LIA) to increase the signal to noise ratio of detected signal. The output signal of the OBD is first sent to one of the lock-in-amplifiers (LIA1) whose reference signal is taken from the modulation frequency of rf signal. The output from LIA1 is then sent to other LIA (LIA2) whose reference signal is taken from the repetition rate of laser beam. To measure the correct phase of time varying sinusoidal spin accumulation signal, the SG is synchronized with the electronic device which triggers the laser pulses and the LIAs are also phase locked. Moreover, we use a green LED to illuminate the device surface with perpendicular incidence through same microscope objective. The back-reflected light is used for real time imaging of device and laser spot with a high spatial resolution charge coupled device (CCD) camera. CCD helps us to place the laser beam at a desired position on the device.

## Results and discussion

Figure 2 shows our GI-XRD measurements for $Cu/Bi_2O_3$ 2(a) and $Ag/Bi_2O_3$ 2(b), at two GI angles equal to 0.30° (black) and 0.40° (red). At 0.30° incidence (black spectra) only one feature is evident, a broad peak at 28° in both interfaces, $Cu/Bi_2O_3$ 2(a) and $Ag/Bi_2O_3$ 2(b). This broad peak is the characteristic peak of our top $Bi_2O_3$ layer, and it indicates a polymorphous structure in $\alpha$-phase, the most stable phase at room temperature of $Bi_2O_3$ [16]. Interestingly, at a GI angle of 0.40°, the Cu and Ag layers at the interface show preferential crystallinity at the interface of the Cu (111) face, same crystallinity orientation is present in the whole thickness of our samples (20 nm). This orientation is similar to previous reports of interfaces with Rashba type spin orbit coupling (SOC) of Ag(111)/Bi [17]. Figure 2(c) shows energy dispersive X-ray (EDX) spectroscopy measurements for $Cu/Bi_2O_3$ (red) and $Ag/Bi_2O_3$ (light blue) in the energy range between 2keV and 9.2keV. We identified peaks corresponding to Bi-M$\alpha$ (2.4 keV), Ag-L$\alpha$ (3.0 keV) and Cu-K$\alpha$ (8.0 keV). Previous report shows similar EDX spectra for Bi-M$\alpha$ peak contained in α-Bi2O3 [18]. Main peak corresponding to Oxygen in metal oxide compounds is out of the range of our apparatus (<1KeV) [19], and usually the weak intensity of EDX for light elements makes their detection challenging. Nevertheless, we can assure the formation and presence of $Bi_2O_3$ compound in $\alpha$-phase by our GI-XRD spectroscopy measurements.

Now, we move on to analyze the breaking of spatial inversion symmetry at our $Cu/Bi_2O_3$ and $Ag/Bi_2O_3$ interfaces. As previously mentioned, the absence of spatial inversion symmetry induces Rashba SOC, which permits spin to charge interconversion experiments [10, 20-22].

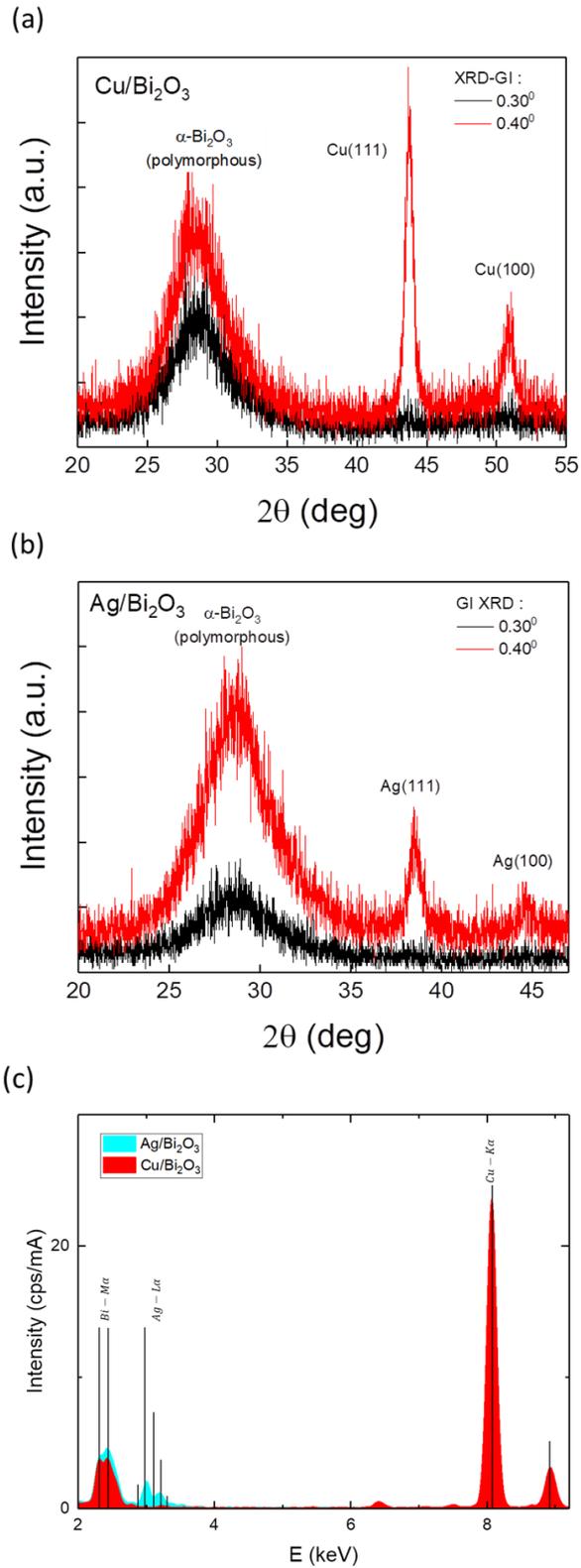

**Figure 2.** GI-XRD patterns, for (a) Cu/Bi$_2$O$_3$ and (b) Ag/Bi$_2$O$_3$ specimens, at 0.30° (black) and 0.40° (red) incidence angles. Energy dispersive X-ray (EDX) spectroscopy measurements (c) for Cu/Bi$_2$O$_3$ (red) and Ag/Bi$_2$O$_3$ (light blue) in the energy range between 2keV and 9.2keV.

For instance, SOC permits to generate spin accumulation by injection of electrical charge current, the so-call direct Rashba - Edelstein effect [11]. For our experiments we inject an AC sinusoidal voltage to our interfaces with a frequency of 100 MHz, which readily induce oscillating spin accumulation. We test the oscillating spin accumulation at three random positions in our interfaces by time resolved transverse magneto optical Kerr effect (TR-TMOKE). Figure 3 and figure 4 show our experimental TR-TMOKE signals for $Cu/Bi_2O_3$ (open circles) and $Ag/Bi_2O_3$ (open squares), respectively.

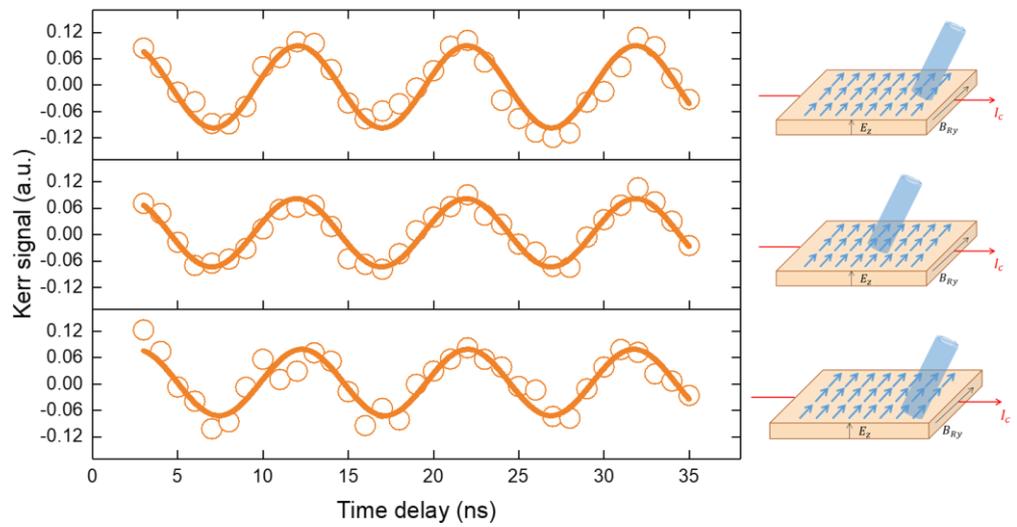

**Figure 3.** Time resolved transverse magneto optical Kerr effect (TR-TMOKE) signal at three different laser beam positions at $Cu/Bi_2O_3$ interface. Laser beam positions are represented in the schematics on the right-hand side.

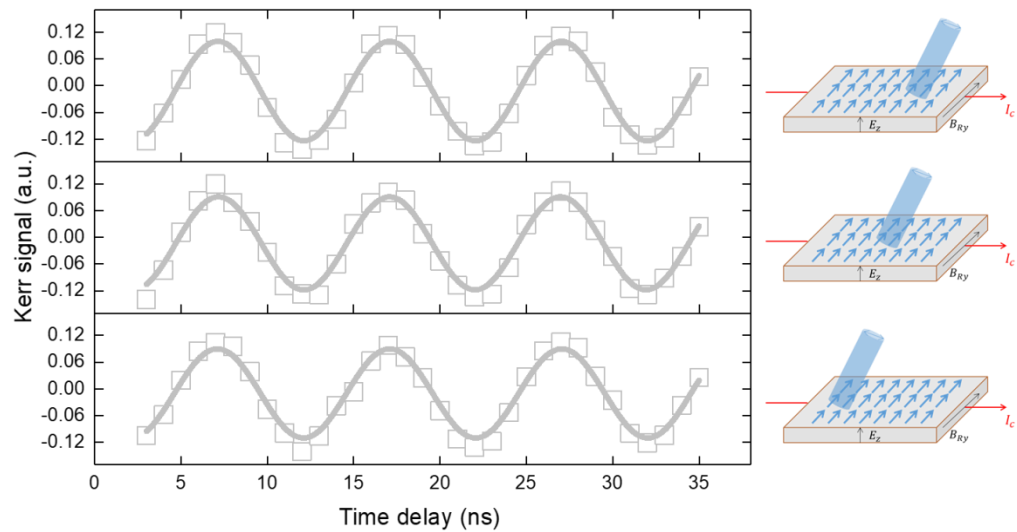

**Figure 4.** Time resolved transverse magneto optical Kerr effect (TR-TMOKE) signal at three different laser beam positions at Ag/Bi$_2$O$_3$ interface. Laser beam positions are represented in the schematics on the right-hand side.

From fitting (solid lines) we can extract a time periodicity of our oscillating signal of 10 ns or 100 MHz in frequency domain, correlated to the frequency of our AC voltage excitation. Moreover, from figure 3 and figure 4 we can observe that TR-TMOKE signals for Cu/Bi$_2$O$_3$ and Ag/Bi$_2$O$_3$ have opposite phases, this implies opposite orientation of spin accumulation. The spin accumulation corresponds to the splitting of the free electron energy dispersion $E_\pm(\mathbf{k}) = E_0 + (\hbar^2 \mathbf{k}^2/2m^*) \pm |\alpha_R||\mathbf{k}|$. The orientation of the spin polarization of the $E_\pm$ is defined by $\mathbf{P}_\pm(\mathbf{k}) = \pm \frac{\alpha_R}{|\alpha_R|}(-k_y, k_x, 0)/|\mathbf{k}|$. Both, the orientation of spin polarization and corresponding spin accumulation are directly dependent on the Rashba parameter $\alpha_R$ and the sign of effective mass $m^*$, which determines the splitting order of inner and outer energy branches ($E_+, E_-$) [23]. The estimated $\alpha_R$ for Cu/Bi$_2$O$_3$ and Ag/Bi$_2$O$_3$ from spin pumping experiments show $\alpha_R = -0.25 \pm 0.03\ eV.\text{Å}$ [20] and $\alpha_R = +0.18 \pm 0.04\ eV.\text{Å}$ [10], respectively, while the sign of effective mass $m^*$ is negative for both interfaces. These characteristics of our measurements correspond to our initial assumption of the existence of Rashba spin orbit interaction at our interfaces. Additionally, spin accumulation induced by direct Rashba - Edelstein effect is expected to be homogenous across the whole interface area. We test this prediction by TR-TMOKE measurements at three random laser spot positions at our interfaces. Similar conclusions to the description above were discussed previously by the authors [10]. Here, we turn our discussion to the estimation of our spin accumulation at Rashba like interfaces, and compare our results to previous reports for spin Hall effect materials, where certain controversy exists.

Intrinsically, spin accumulation generated by bulk spin Hall effect (SHE) and direct Rashba Edelstein effect (DREE) at interfaces have different distribution of spin orientations. Spin accumulation via SHE is oriented perpendicular to the electron flow and opposite at opposite sample planes, see schematic in figure 5(a). In contrast, spin accumulation via DREE at interfaces is in -plane, homogenous and perpendicular to the electron flow direction, figure 5(b). These intrinsic spin orientations are relevant when trying to detect in-plane spin accumulation by optical Kerr / Faraday effects. For instance, if the penetration depth of a given laser in a MOKE setup is larger than half of the sample thickness with SHE, the spin accumulation detected by the laser will have two orientation components related to each half of the sample, resulting in an effective reduction of the total detection of spin accumulation. One way to minimize this issue is to defocus the laser beam, effectively reducing the penetration depth and increasing the collection area at the sample top surface [4].

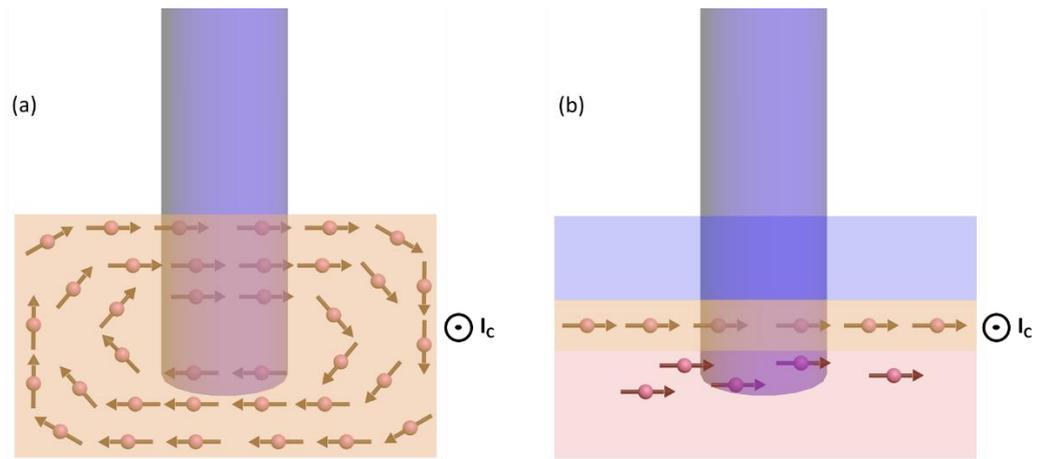

**Figure 5.** Orientation distribution of current induced spin accumulation for bulk spin Hall effect materials (i.e. Pt) (a), and direct Rashba – Edelstein effect at interfaces (i.e. $Cu/Bi_2O_3$) (b).

Similar issue is not present at Rashba interfaces, in fact, the spin accumulation can be enhanced by the coherent tunneling of spins from the interface into the nonmagnetic metal bulk. In our experiments the spin accumulation is generated at the $Cu/Bi_2O_3$ and $Ag/Bi_2O_3$ interface, with a given spin tunnel probability. If the spin tunneling occurs and the spin diffusion length is larger than the thickness of nonmagnetic metal layers, the detected spin accumulation is incremented with the penetration depth of the laser. Recent experiments of spin magneto resistance at the $Cu/Bi_2O_3$ interface observed a diffusion of accumulated spins of 33.9% from the interface into the Cu bulk layer [21]. Additional, at room temperature the spin diffusion length for Cu and Ag is much larger than the thickness in our samples (20nm) [24, 25]. These characteristics indicate that the density of spins detected in our experiments is enhanced by the diffused spins into the bulk, with a detection limit set by the penetration depth of our laser. In our previous report we estimated a total spin accumulation of 14.0 $\mu eV$ and 23.7 $\mu eV$ for $Cu/Bi_2O_3$ and $Ag/Bi_2O_3$ interfaces [10]. The typical thickness of our interfaces is 0.4nm, and the penetration depth of our laser ($\lambda = 408nm$) is about 11 nm for both samples [26]. Therefore, we have an enhancement factor for our detected signal of 9.3, giving an estimated detection of spin accumulation of 130.2 $\mu eV$ and 220.4 $\mu eV$ for Cu/Bi$_2$O$_3$ and Ag/Bi$_2$O$_3$, respectively. These spin accumulation estimations are in good agreement with previous reports of SHE materials with spin accumulation of 180 $\mu eV$ [4, 9].

## Conclusions

In summary, we showed the characterization of the crystal structure at our interfaces by GI-XRD. We observed a broad peak at 28$^0$ at both interfaces, which corresponds to $\alpha$-phase of Bi$_2$O$_3$, its most stable phase at room temperature [16]. Moreover, we observed preferential crystal orientation of the (111) faces for both, Cu and Ag layers. The (111) preferred crystal orientation has been previously observed in other Rashba type interfaces [17]. We showed the optical detection of spin accumulation by Kerr effect. Our signal shows a uniform and preferential spin orientation, which is related to spin accumulation induced by DREE [10, 11]. We further discussed the estimation of our spin accumulation and compare our results with previous reports of MOKE detection of spin accumulation in SHE materials [4, 8, 9]. The present manuscript gives a detail description of our experiments and discusses the feasibility of optical Kerr effect characterization of spin accumulation at interfaces with Rashba type spin orbit interaction. We expect that our work motives further optical exploration of spin accumulation in both Rashba and spin Hall effect systems. In terms of electronic materials, with the slowdown of the complementary metal-oxide-semiconductor (CMOS) technologies, new capabilities based on spin and polarization in metal oxides and ferroelectrics may pave the way forward [28, 29].

**Acknowledgements**

We acknowledge Yoshio Maebashi for technical support. This work was supported by Grant-in-Aid for Scientific Research on Innovative Area, "Nano Spin Conversion Science" (Grant No. 26103002) and RIKEN Incentive Research Project Grant No. FY2016. F.A. was supported by the Ministry of Education, Culture, Sports, Science and Technology (MEXT) Scholarship, Japan

# References

1. Kato, Y.K., Myers, R.C., Gossard, A.C., Awschalom, D.D., Science 306, 1910-1913 (2004)
2. Kurebayashi, H., Sinova, J., Fang, D., Irvine, A.C., Skinner, T.D., Wunderlich, J., Novk, V., Campion, R.P., Gallagher, B.L., Vehstedt, E.K., Zrbo, L.P., Vborn, K., Ferguson, A.J., Jungwirth, T., Nat. Nanotechnology 9, 211217 (2014)
3. Miron, I.M., Garello, K., Gaudin, G., Zermatten, P-J., Costache, M. V., Auffret, S., Bandiera, S., Rodmacq, B., Schuhl, A., Gambardella, P., Nature 476, 189 - 193 (2011)
4. O. M. J. van't Ervea, A. T. Hanbicki, K. M. McCrearyb, C. H. Li, and B. T. Jonker, Appl. Phys. Lett. 104, 172402 (2014)
5. G. M. Choi, B. C. Min, K. J. Lee, and D. G. Cahill, Nat. Commun. 5, 4334 (2014).
6. Y. Su, H. Wang, J. Li, C. Tian, R. Wu, X. Jin, and Y. R. Shen, Appl. Phys. Lett. 110, 042401 (2017)
7. P. Riego, S. Velez, J. M. Gomez-Perez, J. A. Arregi, L. E. Hueso, F. Casanova, and A. Berger, Appl. Phys. Lett. 109, 172402 (2016)
8. C. Stamm, C. Murer, M. Berritta, J. Feng, M. Gabureac, P. M. Oppeneer, and P. Gambardella, Phys. Rev. Lett. 119, 087203 (2017)
9. O. M. J. van't Erve, A. T. Hanbicki, K. M. McCreary, C. H. Li, and B. T. Jonker, preprint arXiv:1703.03844 [cond-mat.mes-hall] (2017)
10. J. Puebla, F. Auvray, M. Xu, B. Rana, A. Albouy, H. Tsai, K. Kondou, G. Tatara, and Y. Otani, Appl. Phys. Lett. 111, 092402 (2017)
11. P. Gambardella and I. M. Miron, Philos. Trans. R. Soc. A 369, 3175–3197 (2011).
12. B. Rana, Y. Fukuma, K. Miura, H. Takahashi, and Y. Otani, Appl. Phys. Lett. 111, 052404 (2017).
13. A. Barman, T. Kimura, Y. Otani, Y. Fukuma, K. Akahane, and S. Meguro, Rev. Sci. Instrum. 79, 123905 (2008).
14. D. A. Allwood, P. R. Seem, S. Basu, P. W. Fry, U. J. Gibson, and R. P. Cowburn, Appl. Phys. Lett. 92, 072503 (2008)
15. Hideo Kawaguchi and Gen Tatara, Phys. Rev. B 94, 235148 (2016)

16. A. Walsh, G. W. Watson, D. J. Payne, R. G. Edgell, J. Guo, P-A. Glans, T. Learmonth, and K. E. Smith Phys. Rev. B 73, 235104 (2006)
17. J. C. Rojas-Sanchez, L. Vila, G. Desfonds, S. Gambarelli, J. P. Attane, J. M. De Teresa, C. Magen, and A. Fert, Nat. Commun. 4, 2944 (2013)
18. Y. Astuti, A. Fauziyah, S. Nurhayati, A.D. Wulansari, R. Andianingrum, A.R. Hakim and G. Bhaduri, IOP Conf. Ser.: Mater. Sci. Eng. 107 012006 (2016)
19. A. A. Yadav, A. C. Lokhande, P. A. Shinde, J. H. KimEmail and C. D. Lokhande, J Mater Sci: Mater Electron, 28: 13112 (2017)
20. S. Karube, K. Kondou, Y. Otani, Applied Physics Express 9, 033001 (2016)
21. J. Kim, Y-T. Chen, S. Karube, S. Takahashi, K. Kondou, G. Tatara, and Y. Otani, Phys. Rev. B 96, 140409(R) (2017)
22. H. Tsai, S. Karube, K. Kondou, N. Yamaguchi, F. Ishii and Y. Otani, Scientific Reports 8, 5564 (2018)
23. H. Bentmann, T. Kuzumaki, G. Bihlmayer, S. Blugel, E.V. Chulkov, F. Reinert, K. Sakamoto, Phys. Rev. B 84, 115426 (2011)
24. T. Kimura, J. Hamrle, and Y. Otani, Phys. Rev. B 72, 014461 (2005)
25. T. Kimura and Y. Otani, Phys. Rev. Lett. 99, 196604 (2007)
26. P. B. Johnson and R. W. Christy, Phys. Rev. B 6, 4370 (1972)
27. A. Walsh, G. W. Watson, D. J. Payne, R. G. Edgell, J. Guo, P-A. Glans, T. Learmonth, and K. E. Smith Phys. Rev. B 73, 235104
28. S. Manipatruni, D. E. Nikonov and I. A. Young, Nat. Phys. 14, 338-343 (2018)
29. J. Varignon, L. Vila, A. Barthelemy and M. Bibes, Nat. Phys. 14, 322-325 (2018)